\documentclass[aps,prd,showpacs,reprint]{revtex4-1}

\usepackage[utf8]{inputenc}
\usepackage{bm}
\usepackage{ae,aecompl}
\usepackage{graphicx}
\usepackage{amsmath}
\usepackage{amssymb}
\usepackage[T1]{fontenc}
\newcommand{\vb}[1]{\bm{ #1}} % type setting for bold vector
\newcommand{\unite}[1]{\,\mathrm{#1}}
\newcommand{\op}[1]{\mathrm{#1}}
\newcommand{\tv}[1]{\vec{ #1}} % type setting of a three-vector
\newcommand{\gw}{gravitational wave}
\newcommand{\gws}{gravitational waves}
\newcommand{\Gw}{Gravitational wave}

\def\xp{{\textsc{X-Pipeline}}}

\begin{document}

\title[Externally triggered search performance ]{Performance of
an externally triggered gravitational-wave burst search}

\author{Michał~Wąs}
\email{michal.was@aei.mpg.de}
\affiliation{LAL, Université Paris-Sud, IN2P3/CNRS, F-91898 Orsay, France}
\affiliation{Albert-Einstein-Institut, Max-Planck-Institut für
  Gravitationsphysik, D-30167 Hannover, Germany}

\author{Patrick~J.~Sutton, Gareth~Jones}
\affiliation{Cardiff University, Cardiff, CF24 3AA, United Kingdom}

\author{Isabel Leonor}
\affiliation{University of Oregon, Eugene, OR 97403, USA}

\begin{abstract}
  We present the performance of searches for 
  gravitational wave bursts
  associated with external astrophysical triggers as a function of
  the search sky region. We discuss both the case of Gaussian noise
  and real noise of gravitational wave detectors for arbitrary
  detector networks. We demonstrate the ability to reach Gaussian
  limited sensitivity in real non-Gaussian data, and show the
  conditions required to attain it. We find that a single sky
  position search is $\sim$20\% more sensitive than an
  all-sky search of the same data.
\end{abstract}

\pacs{04.80.Nm, 07.05.Kf}
\maketitle

\section{Introduction}

Searches for transient gravitational waves (bursts) typically fall
into one of two categories: all-sky untriggered searches, which scan
the entire sky and search throughout the available data; and triggered or
directed searches, which historically analyze only a single point on
the sky corresponding to some astrophysical source of interest 
(see for example Refs.~\cite{burstS5y2,burstGrbS5}). 
However, some instruments provide only an approximate sky
location for external triggers, which requires scanning a relatively
large patch on the sky for an associated gravitational wave. 
For instance, this is the case for gamma-ray
bursts (GRBs) localized by the Gamma-ray Burst Monitor (GBM) 
\cite{Briggs09, GBMsysErr_GCN} on Fermi, and 
for some high energy neutrino candidates~\cite{Baret:2011nu,2012arXiv1205.3018T}.

We present an implementation of a gravitational wave burst (GWB) search
which is able to scan arbitrary sky patches, and demonstrate how the
sensitivity of the analysis varies with the sky region searched.  
In Gaussian noise we find that the sensitivity is a
function of the total signal-to-noise-ratio received by the
gravitational wave detector network, which is the expected result for
an optimal search in Gaussian noise. However for real non-Gaussian 
gravitational wave detector noise, we find a different dependence of
the sensitivity, due to the requirement that the gravitational wave 
signal be seen by
at least two detectors to be distinguished from spurious noise
transients. We obtain an empirical formula for the sensitivity of the
search in real noise as a function of the sky position for arbitrary
detector networks.
In particular, a sensitivity loss of $\sim$20\% is observed between searching at a
single sky position and an all-sky search of the same data.

Finally, we demonstrate the ability to reach Gaussian limited sensitivity in
real non-Gaussian data, and show the conditions required to attain it.
Specifically, we find that all of the detectors present 
in the network need to have comparable sensitivity. Moreover,
we show that in practice adding a detector to a gravitational wave
detector network may actually {\em reduce} the search sensitivity for some sky
areas when analyzing real data.

We begin in Sec.~\ref{sec:overview} with a brief introduction to
coherent searches for GWBs. We follow in Sec.~\ref{sec:detStat}
with details of the detection statistic and sky scanning algorithm
used in this paper.  The spurious-noise rejection tests are presented
in Sec.~\ref{sec:cohCuts}. In Sec.~\ref{sec:performance} we
present the performance of the search for GWBs; in particular, the sky
dependence is discussed in Sec.~\ref{sec:skyDep}, the comparison
between real and Gaussian noise is shown in
Sec.~\ref{sec:sensDistrib}, and the dependence on the size of the
sky region is presented in Sec.~\ref{sec:areaDep}. We conclude with 
some comments on the implications of these results in Sec.~\ref{sec:conclusion}.

\section{Coherent analysis overview}\label{sec:overview}

Coherent analysis of \gw\ data was originally introduced
in Ref.~\cite{Gursel89}. Since then it has proven to be an effective
method for GWB searches in LIGO-Virgo data, and it is now the dominant search
methodology~\cite{burstS5y2,burstGrbS5,S4LIGOGEO,burstS5y1,S5IMBBH,burstS6allsky,grb051103,grbS6,2012arXiv1205.3018T}.
Here we give a short overview of coherent
analysis in order to introduce notation used in the following sections.

For a gravitational wave $\left(h_+(t),h_\times(t)\right)$ incoming
from a sky location $\hat{\Omega}$ the calibrated data from a
gravitational wave detector $\alpha$ are of the form
\begin{multline}
  \label{eq:data}
  d_\alpha(t+\Delta t_\alpha(\hat\Omega))
		 =   \\F_\alpha^+(\hat\Omega) h_+(t) + F_\alpha^\times(\hat\Omega) h_\times(t)
                 + n_\alpha(t+\Delta t_\alpha(\hat\Omega)) \,.
\end{multline}
Here $F_\alpha^+(\hat\Omega)$, $F_\alpha^\times(\hat\Omega)$ are the
\emph{antenna response functions} of the given detector to the plus ($+$) and
cross ($\times$) polarized gravitational waves, $n_\alpha$ is a time series of 
detector noise, and $\Delta t_\alpha(\hat\Omega)$ is the
gravitational wave travel time between the position $\tv{r}_\alpha$ of
the detector and an arbitrary reference point $\tv{r}_0$:
\begin{equation}
  \label{eq:delay}
  \Delta t_\alpha(\hat\Omega) = \frac{1}{c}(\tv{r}_0-\tv{r}_\alpha)\cdot\hat\Omega \,.
\end{equation}

For the case where the sky location $\hat{\Omega}$ is known {\it a priori}, 
the first step of the analysis is to time shift the data $d_\alpha$
by the known $\Delta t_\alpha(\hat\Omega)$ in order to obtain a
gravitational wave contribution which is synchronous between the
different data time series. 
The data from each detector are then whitened and decomposed in a
time-frequency representation, e.g. using a short Fourier transform 
or a wavelet decomposition, where the time-frequency basis functions 
typically have length between several milliseconds and several hundred 
milliseconds. For a given time-frequency pixel (basis function) of center time
$t$ and center frequency $f$ the obtained decomposition can be written
compactly as 
\begin{equation}
  \vb{d} = \vb{F}^+ h_+ + \vb{F}^\times h_\times
  + \vb{n} \,,
\end{equation}
where boldface symbols denote the vector of whitened,
time-frequency decomposed time series in the $D$ dimensional
space of detectors:
\begin{subequations}
  \label{eq:boldVect}
  \begin{align}
  &\vb{d} =
  \begin{bmatrix}
    d_1(t,f)/\sqrt{S_1(f)} \\ \vdots \\ d_D(t,f)/\sqrt{S_D(f)}
  \end{bmatrix}\\
  &\vb{n} =
  \begin{bmatrix}
    n_1(t,f)/\sqrt{S_1(f)} \\ \vdots \\ n_D(t,f)/\sqrt{S_D(f)}
  \end{bmatrix}\\
  &\vb{F}^{+,\times} =
  \begin{bmatrix}
    F^{+,\times}_1/\sqrt{S_1(f)} \\ \vdots \\ F^{+,\times}_D/\sqrt{S_D(f)}
  \end{bmatrix}    
\end{align}
\end{subequations}
Here $S_i(f)$ is the one-sided noise power spectrum of detector $i$.
Note that the gravitational wave contributions $h_+$, $h_\times$ are
the projection on a time-frequency basis function without any
whitening.

The basis used in Eqs.~\eqref{eq:boldVect} to describe
vectors in the $D$ dimensional space of detectors is not adapted to
the gravitational wave contribution. Given that the data are whitened
and detector noise can be assumed to be uncorrelated between detectors,
the vector $\vb{n}$ has an identity covariance matrix, which is
invariant under change of orthonormal basis. Hence we can construct a new
adapted orthonormal basis, in which the first two vectors span the
gravitational wave plane~\footnote{For simplicity we consider only the 
  case of a network composed of non-aligned detectors, as in the  
  most recent science run of the LIGO-Virgo detector network 
  (only one gravitational wave detector was operational at the 
  LIGO-Hanford site during the 2009-2010 
  run).} generated by the $\vb{F}^+$ and $\vb{F}^\times$
vectors, and the $D-2$ remaining vectors span the \emph{null space},
the space orthogonal to the gravitational wave plane. 

This basis can be further refined in various ways, such as by choosing 
the first two vectors along the directions of maximal and minimal response 
to a linearly polarized gravitational wave. These two directions are 
orthogonal and correspond to the \emph{dominant polarization} 
choice~\cite{Klimenko05} of the arbitrary
gravitational wave polarization angle reference, which is a particular
choice of the definition of the plus and cross polarizations. We denote
the antenna response vectors for this special polarization choice by
$\vb{f}^+$ and $\vb{f}^\times$. They have the properties
\begin{subequations}
  \begin{align}
    & |\vb{f}^+|^2 \geq |\vb{f}^\times|^2\,, \\
    & \vb{f}^+ \cdot \vb{f}^\times = 0 \,.
  \end{align}
\end{subequations}
Note that the choice of which of the plus and cross vector has a
larger amplitude is purely conventional.
The unit vectors of our adapted basis are hence $\vb{e}^+ =
\vb{f}^+/|\vb{f}^+|$ and $\vb{e}^\times =
\vb{f}^\times/|\vb{f}^\times|$, complemented by vectors spanning the
null space, for instance $\vb{e}^n = \vb{e}^+ \wedge \vb{e}^\times$ for
the case of 3 non-aligned detectors.

An alternative basis choice may be appropriate when we have prior 
information on the expected gravitational wave polarization.  
For example, for a circularly polarized \gw\ signal the projection onto a
time-frequency basis function of the two polarizations are related by
\begin{equation}
  h_\times = \pm i h_+\,,
\end{equation}
depending on whether the signal is left or right-hand polarized. Hence
in the detector space the \gw\ contribution will lie along either the
left or right-handed response vectors
\begin{equation}
  \vb{f}^\circlearrowright = \vb{f}^+ + i \vb{f}^\times, \qquad \vb{f}^\circlearrowleft = \vb{f}^+ - i
\vb{f}^\times \, ,
\end{equation}
with corresponding unit vectors $\vb{e}^{\circlearrowright} =
\vb{f}^{\circlearrowright}/|\vb{f}^{\circlearrowright}|$,
$\vb{e}^{\circlearrowleft} =
\vb{f}^{\circlearrowleft}/|\vb{f}^{\circlearrowleft}|$. To construct a
left or right-handed basis we use the vectors orthogonal to the
response vectors in the \gw\ plane
\begin{equation}
      \vb{f}^{n\circlearrowright} =
    \frac{\vb{f}^+}{|\vb{f}^+|^2}
    -i\frac{\vb{f}^\times}{|\vb{f}^\times|^2}\,,  \quad
    \vb{f}^{n\circlearrowleft} =
    \frac{\vb{f}^+}{|\vb{f}^+|^2}
    +i\frac{\vb{f}^\times}{|\vb{f}^\times|^2}\,,
\end{equation}
with the corresponding unit vectors $\vb{e}^{n\circlearrowright} =
\vb{f}^{n\circlearrowright}/|\vb{f}^{n\circlearrowright}|$ and
$\vb{e}^{n\circlearrowleft} =
\vb{f}^{n\circlearrowleft}/|\vb{f}^{n\circlearrowleft}|$. 

Either basis (dominant or circular) can be used to construct two types 
of statistics: {\it detection statistics} and {\it coherent consistency 
statistics}. A detection
statistic is used to ranks events as more consistent with a given
model of signal buried in noise than a model of noise only. It is
usually expressed as the log-likelihood ratio between these two
models. Coherent consistency tests, on the other hand, are used to reject spurious
noise transients which are usually not included in the noise model of
the detection statistic. We discuss the formulation of detection 
statistics in Sec.~\ref{sec:detStat}, and coherent consistency tests 
in Sec.~\ref{sec:cohCuts}; for the moment we note that 
these statistics are only functions of the data vector $\vb{d}$ and the
antenna response vectors $\vb{f}^+$ and $\vb{f}^\times$ or $\vb{f}^{\circlearrowright},\vb{f}^{\circlearrowleft}$ 
and $\vb{f}^{n\circlearrowright},\vb{f}^{n\circlearrowleft}$. The value of each statistic is computed
independently for each time-frequency pixel, and the resulting values
of the detection statistic over the array of time-frequency pixels is
used to define \gw\ events. Specifically, all time-frequency pixels with
detection statistic above a certain threshold are clustered to form
events, for instance using nearest-neighbor
clustering~\cite{Sylvestre02}. The final detection statistic of such an
event is simply the sum of the detection statistic of all the
pixels composing the event due to the additive properties of the
log-likelihood ratio. The other statistics are also summed over the
cluster of pixels composing the event.

This event generation procedure is used on numerous background samples
(generated from real data using the \emph{time slide} technique) and signal
samples (generated by adding simulated gravitational wave signals to
real data). The obtained background and signal events are
used to tune the coherent consistency tests to reject the tail of
spurious noise transients inconsistent with the \gw\ signal hypothesis. 
Independent samples of background events are then used to estimate
the distribution of background events that survive the consistency tests, 
which is used in turn to define the statistical significance of 
any candidate \gw\ events from the analysis of data coincident with 
the external astrophysical trigger. An independent
sample of simulated signal events is used to estimate the sensitivity of the
analysis as a function of \gw\ signal amplitude, and to construct upper
limits on the \gw\ signal amplitude whenever no 
significant \gw\ event is found.

\section{Detection statistic}\label{sec:detStat}

In a \gw\ search, the detection
statistic is used to ranks events as more consistent with a given
model of signal buried in noise than with a model of noise only.
The detection statistic is often based on some measure of the 
energy in the data, motivated by a likelihood-ratio analysis.
For the present analysis, we construct a detection statistic following the Bayesian formalism
of Ref.~\cite{Searle08}. The signal model is a circularly polarized \gw\
signal with Gaussian amplitude distribution of width $\sigma_h$, 
and the noise model is Gaussian. The circular polarization assumption
is well motivated for some astrophysical sources, for instance
gamma-ray bursts, as discussed in Sec.~\ref{sec:circPolCuts}. For a
right circularly polarized signal we obtain the log-likelihood ratio
\begin{equation}
  2L(\vb{d}|\circlearrowright,\sigma_h) = \frac{|\vb{e}^\circlearrowright \cdot \vb{d}|^2}{1+1/(\sigma_h |\vb{f}^\circlearrowright|)^2} -
  \log(1 + \sigma_h^2 |\vb{f}^\circlearrowright|^2)\,,
\end{equation}
and the left circular polarization log-likelihood ratio
$L(\vb{d}|\circlearrowleft,\sigma_h)$ has an analogous form. The final
likelihood ratio is obtained by marginalizing over the left versus
right choice, and over a discrete set $\mathcal{A}$ of $\sigma_h$ covering the range
$[10^{-23}, 10^{-21}]\unite{Hz^{1/2}}$ of realistic detectable
signals. The detection statistic used thus has the form
\begin{multline}\label{eq:Sdet}
  S_\text{detection} = L(\vb{d}) = \\\log \sum_{\sigma_h \in \mathcal{A}}
  \frac{1}{2|\mathcal{A}|} \left[\exp L(\vb{d}|\circlearrowright,\sigma_h) +
  \exp L(\vb{d}|\circlearrowleft,\sigma_h)\right].
\end{multline}

The discussion so far has assumed we know the sky position $\hat{\Omega}$ 
of the \gw\ source {\it a priori}.  For some searches this is indeed the case, 
such as for gamma-ray bursts detected by the Swift satellite~\cite{BAT05}.
In other cases, such as untriggered all-sky searches 
\cite{burstS5y2,S4LIGOGEO,burstS5y1,S5IMBBH,burstS6allsky}, $\hat{\Omega}$ is 
not known, or may only be constrained to some large region of the sky.  
An example of the latter is gamma-ray bursts detected by the 
GBM, which has relatively large sky location systematic uncertainties 
of a few
degrees~\cite{Briggs09} and statistical errors 
%typically between a fraction of and a 
of up to $\sim$10 degrees depending on the $\gamma$-ray flux and
spectrum. An error of this size in the sky location used to
synchronize the data time series from \gw\ detectors causes 
timing discrepancies of up to several milliseconds. A potential \gw\ signal at a few
hundred Hertz could easily be shifted by a quarter of a period or more between a
pair of \gw\ detectors, and the signal could be rejected by a coherent
consistency test.

The standard solution in \gw\ coherent searches is to repeat the
analysis over a discrete grid of sky positions covering most of the
source sky location probability distribution. Here we use a simple
regular grid, composed of concentric circles around the best estimate
of the source sky location, which covers at least 95\% of the sky
location probability distribution. The constant grid step is chosen so
that the timing synchronization error between any sky location in the
error box and the nearest analysis grid point is less than 10\% of the
period for the highest frequency \gw\ signals included in the search.
This is small enough to limit the amplitude SNR loss due to timing 
error to be less than 10\%.

\Gw\ triggers are produced independently for each sky position grid
point. The detection statistic $S_\text{detection}$, written as a 
log-likelihood ratio between a signal and a background model, 
is penalized by the probability $p_\text{EM}(\Omega)$ of the trial sky
location being the true one, by adding the logarithm of
that probability to the detection statistic
\begin{equation}
  S_\text{penalized} = S_\text{detection} + \log p_\text{EM}(\Omega)\,.
\end{equation}
The reconstructed source sky position for a given signal is the sky
position for which the trigger has the largest penalized detection
statistic. Only that maximal trigger is kept by the analysis of the grid
of sky positions.

As a simple model of $\gamma$-ray satellite errors we use a Fisher
probability~\cite{Briggs99} 
\begin{equation}
p_\text{EM}(\Omega) =   p_\text{Fisher}(\theta;\kappa) =  \frac{\kappa \sin \theta}{\op{e}^\kappa - \op{e}^{-\kappa}}
  \op{e}^{\kappa \cos \theta},
\end{equation}
where $\theta$ is the angle between the best estimate sky location and
the analyzed one.  The parameter $\kappa$ is chosen so that the
95\% coverage radius of this Fisher distribution is equal to the 95\%
coverage radius for a given GRB sky location reconstruction
(with statistical and systematic errors added in quadrature).
This model is a reasonable approximation for localization performed by a single
$\gamma$-ray spacecraft, such as Fermi or Swift, however it may not
apply to other instruments, for instance to localization by the Third
Interplanetary Network of satellites~\cite{IPN09}.

\section{Coherent consistency tests}\label{sec:cohCuts}

\subsection{General framework}

Detection statistics such as Eq.~\eqref{eq:Sdet} are usually
constructed assuming the background detector noise is Gaussian. 
These statistics do not take into account spurious noise
transients, mainly because no good model of these transients is
available.  However these transients are uncorrelated between the
different \gw\ detectors, and powerful coherent consistency tests to 
reject them can be constructed on that basis~\cite{Cadonati04,Wen05,Chatterji06,Klimenko08}. 

An effective method for rejecting noise transients is to project the
\gw\ data vector $\vb{d}$ onto the null space, and to compare the
squared magnitude of this projection with the autocorrelation terms of
that projection~\cite{Chatterji06}. For simplicity let us consider the
case of a one dimensional null space along a vector $\vb{e}^n$. The
squared magnitude of the projection on this vector, also called the 
\emph{coherent null energy}, is
\begin{equation}
  E_n = |\vb{e}^n \cdot \vb{d}|^2 = \sum_{\alpha,\beta} {e^{n*}_\alpha}
  e^n_\beta \, {d_\alpha^*} d_\beta^{\vphantom{*}}\,.
\end{equation}
The autocorrelation part of the null energy, called the 
\emph{incoherent null energy}, is
\begin{equation}
  I_n = \sum_{\alpha} {e^{n*}_\alpha}
  e^n_\alpha \, {d_\alpha^*} d_\alpha^{\vphantom{*}} = \sum_{\alpha} |e^n_\alpha|^2 |d_\alpha^{\vphantom{*}}|^2\,.
\end{equation}
For a strong \gw\ signal the contribution from the different detectors
cancel each other in the coherent null energy, but stay present in the
incoherent null energy, hence $E_n \ll I_n$ is expected. For
noise transients the contributions of each detector are expected to be
uncorrelated, hence $E_n \simeq I_n$. Thus, a threshold on the ratio of the
incoherent and coherent null energy can be used to reject noise
transients.

Several extensions of this framework have been previously implemented
and discussed in Ref.~\cite{Sutton10}. 
First, the separation between the coherent and incoherent energy for both noise
transients and \gw\ signals depends on the value of the incoherent
energy, that is the strength of the deviation from the Gaussian noise
hypothesis, hence a more complicated separation line in the
coherent/incoherent energy plane than a simple ratio is used in practice.
Second, the framework has also been extended to projection vectors that are
not in the null space, but in the gravitational wave plane. The
incoherent energy will remain comparable to the coherent energy for
the case of noise transients, but for gravitational wave signals it
will be much smaller or larger than the coherent energy. The
previously proposed extension~\cite{Sutton10} is to use the plus and cross
polarization directions in the dominant polarization frame. Along
$\vb{e}^+$ a coherent buildup of energy is expected for most 
gravitational wave signals; this projection corresponds to the hard
constraint introduced in Ref.~\cite{Klimenko05}. On the other hand, for many network
configurations and large fractions of the sky $|\vb{f}^\times| \ll 
|\vb{f}^+|$~\cite{Klimenko05}, and the projection on the $\vb{e}^\times$ vector can be
considered as an effective null stream.

\subsection{Extension to circular polarization}\label{sec:circPolCuts}

The main issue with the previously described projections is that none
of them is effective for the case of two \gw\ detectors which see
roughly independent linear polarizations~\footnote{
  This happens when the orientation of the detector arms differs 
  by $45^\circ$ when projected onto the plane orthogonal to the \gw\
  direction of propagation.},
which occurs frequently for networks consisting of one LIGO 
detector plus Virgo. In that case the coherent consistency tests
based on the plus and cross energies perform poorly, as the
cross-correlation terms are small, and the incoherent and coherent
energy are roughly equal for both gravitational waves and noise transients.
This issue was noted in the all-sky GWB search of
2005-2007 LIGO-Virgo data~\cite{burstS5y2}, and can also be seen in
the poor upper limits for the search in association with GRBs of the
same data~\cite{burstGrbS5}.

Here we propose to expand the possible projection vectors
using the assumption of a circularly polarized \gw\ signal. This
imposes correlations between otherwise independent linear
polarizations, and allows us to construct effective consistency tests even for
the case of two strongly misaligned \gw\ detectors. Also, we note that the circular
polarization assumption is well motivated for many astrophysical scenarios. For
instance, for gamma-ray bursts the expected beamed
emission of the gamma rays means that the progenitor is seen roughly
along its axis of rotation. Furthermore, gravitational wave emission models
which could be seen at extra-galactic distances emit predominantly
circularly polarized gravitational waves along their rotation
axis~\cite{Kobayashi03-1,Shibata03,Davies02,Piro07,Romero10}.

We consider the projection onto the manifold of circularly polarized
\gws, which is formed by the two
complex lines along the left and right handed polarization 
directions. The magnitude of the projection is
simply the maximum of the projection on the left and right handed
response unit vectors $\vb{e}^\circlearrowright$ and
$\vb{e}^\circlearrowleft$, which we defined in Sec.~\ref{sec:overview},
\begin{equation}
  E_\text{circular} = \max\left( |\vb{e}^\circlearrowright \cdot \vb{d}|^2,
  |\vb{e}^\circlearrowleft \cdot \vb{d}|^2 \right)\,.
\end{equation}
This projection can be compared to its incoherent part
\begin{equation}
  I_\text{circular} = I_\circlearrowright = I_\circlearrowleft\,,
\end{equation}
which is equal for the right and left
projections. Using the pair of incoherent/coherent energies
$E_\text{circular}, I_\text{circular}$ a consistency test can be
constructed analogously to the ones based on the plus or cross
energies. 

Furthermore, given the circular polarization assumption a null test
can be constructed by considering the unit vectors
$\vb{e}^{n\circlearrowright}$ and $\vb{e}^{n\circlearrowleft}$ that
are orthogonal to the circular projection and defined in
Sec.~\ref{sec:overview}. The null circular energy is defined as
the minimum of the magnitude of the projection onto these two unit
vectors
\begin{equation}
  E_\text{null circular} = \min\left(|\vb{e}^{n\circlearrowright} \cdot \vb{d}|^2,
    |\vb{e}^{n\circlearrowleft} \cdot \vb{d}|^2\right).
\end{equation}
As in the previous cases an incoherent counterpart can be
defined, with the autocorrelation terms of the null left and right
projection being equal.  A consistency test can then be constructed 
using $E_\text{null circular}$, $I_\text{null circular}$.

\section{Analysis performance}\label{sec:performance}

\begin{figure}
  \centering
  \includegraphics[width=\linewidth]{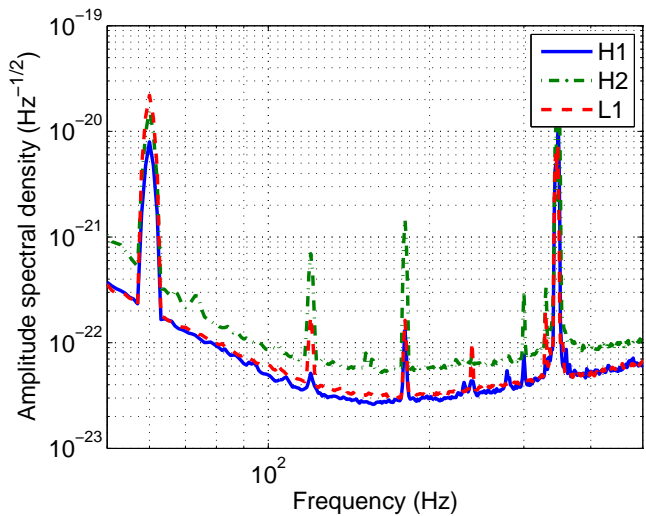}
  \caption{Strain noise spectra from the LIGO detectors used in
    this study. Note that H2 data are treated as if the detector was
    located at the Virgo site in order to simulate the \gw\ detector
    network operated in 2009-2010.}
  \label{fig:itfSpectra}
\end{figure}

The analysis methodology described above has been implemented in 
\xp~\cite{Sutton10}, a software package designed for GWB searches in
association with external astrophysical triggers. Earlier versions of
this package have been used to search for GWBs associated with
GRBs~\cite{burstGrbS5,grb051103} . 

In order to understand the sky dependence of the sensitivity of GWB 
searches, we study the performance of \xp\ in searching for circularly
polarized \gw\ bursts.  Specifically, we characterize the performance for a 
typical GRB-trigger scenario for the most recent science run of the 
LIGO-Virgo network~\cite{iLIGO,Virgo,Virgo11}, from 2009-2010.  Since data from that period have  
not yet been released for data analysis performance studies of the
type presented here, we use as a proxy a 3 hour long sample of LIGO 
data from 23 February 2006. At that time the two detectors at the Hanford site
and the detector at the Livingston site were taking science quality
data; the spectra of the data at the center of that sample are shown on
Fig.~\ref{fig:itfSpectra}.  To be representative of the full network
of large scale interferometric \gw\ detectors which was operational
during 2009-2010, we use the data from 2 km detector Hanford site as if
that detector was located at the Virgo site, 
hence throughout this article we will denote 
these data as from V1. The factor $\sim$2 difference in sensitivity is
roughly representative in the difference in sensitivity between Virgo
and the two 4 km LIGO detectors (H1 and L1) in the 2009-2010 data set.

These data are used to generate background data samples using the time
slide method, and to generate simulated signal samples by adding circularly
polarized sine-Gaussian and compact binary inspiral waveforms into
these data. We perform full autonomous analyses as used in real
externally triggered \gw\ searches to determine the sensitivity for these
signal models. We use a 660\,s long time window around fiducial
external triggers, and search for gravitational waves over the frequency band  
64$-$500 Hz. These are the parameters used in the search for gravitational waves
associated with GRBs in 2009-2010 LIGO-Virgo data~\cite{grbS6}. For
reference, we also use simulated Gaussian noise with the same spectral
properties. With this reference the effect of non-Gaussian noise
transients present in real data can be assessed.

\subsection{Single sky position analysis}\label{sec:skyDep}

\begin{figure}
  \centering
  \includegraphics[width=\linewidth]{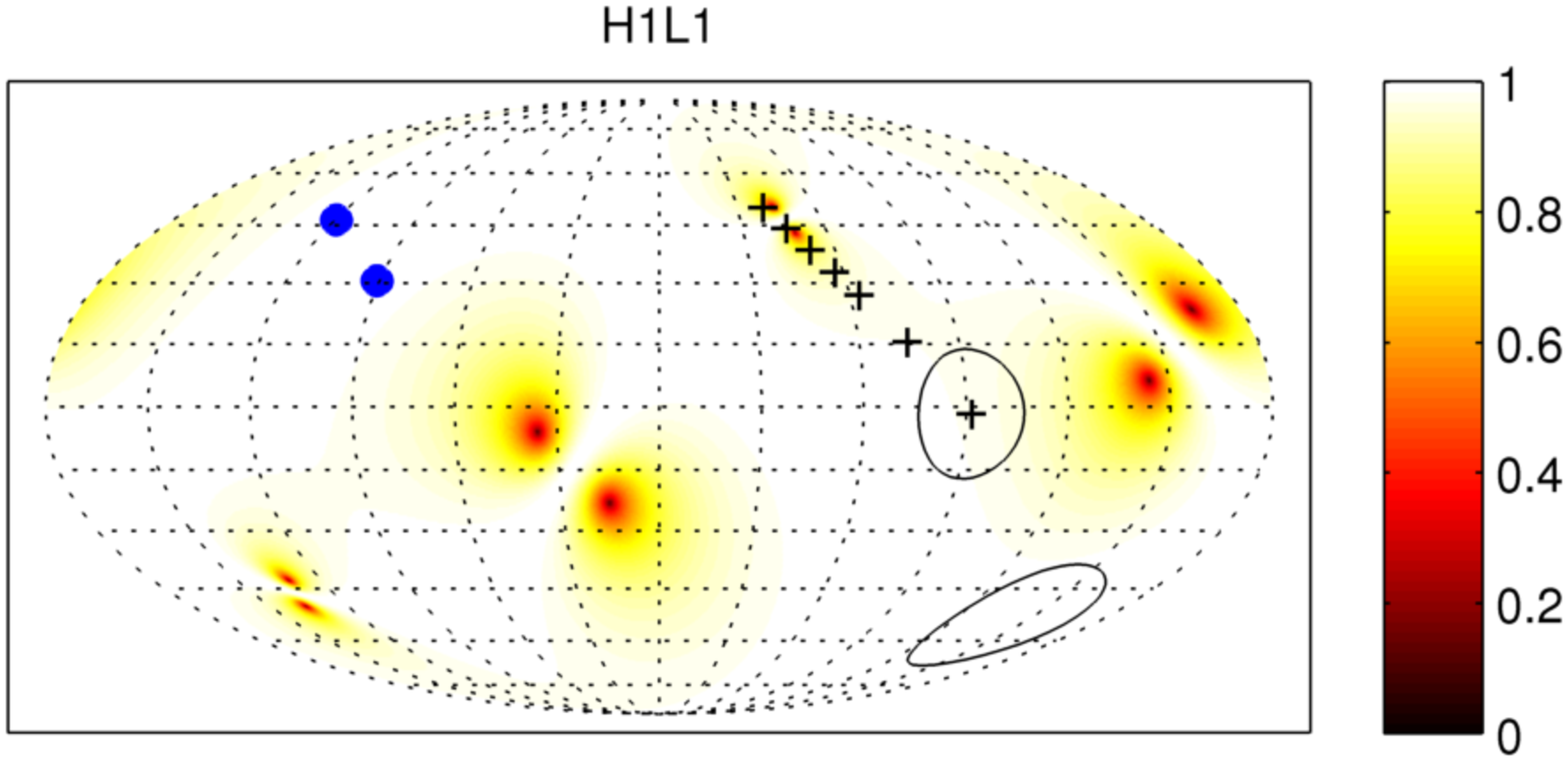}
  \includegraphics[width=\linewidth]{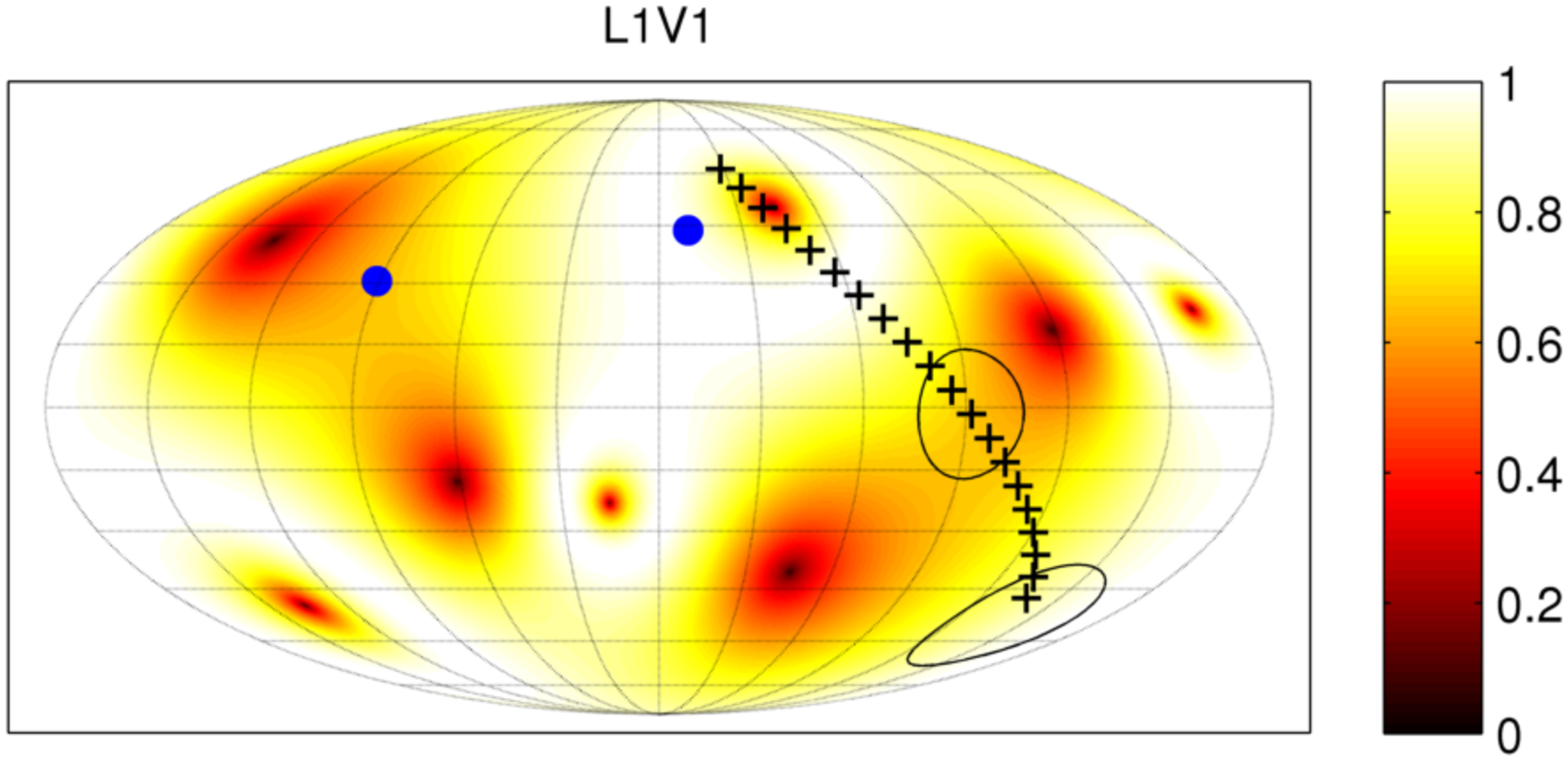}
  \includegraphics[width=\linewidth]{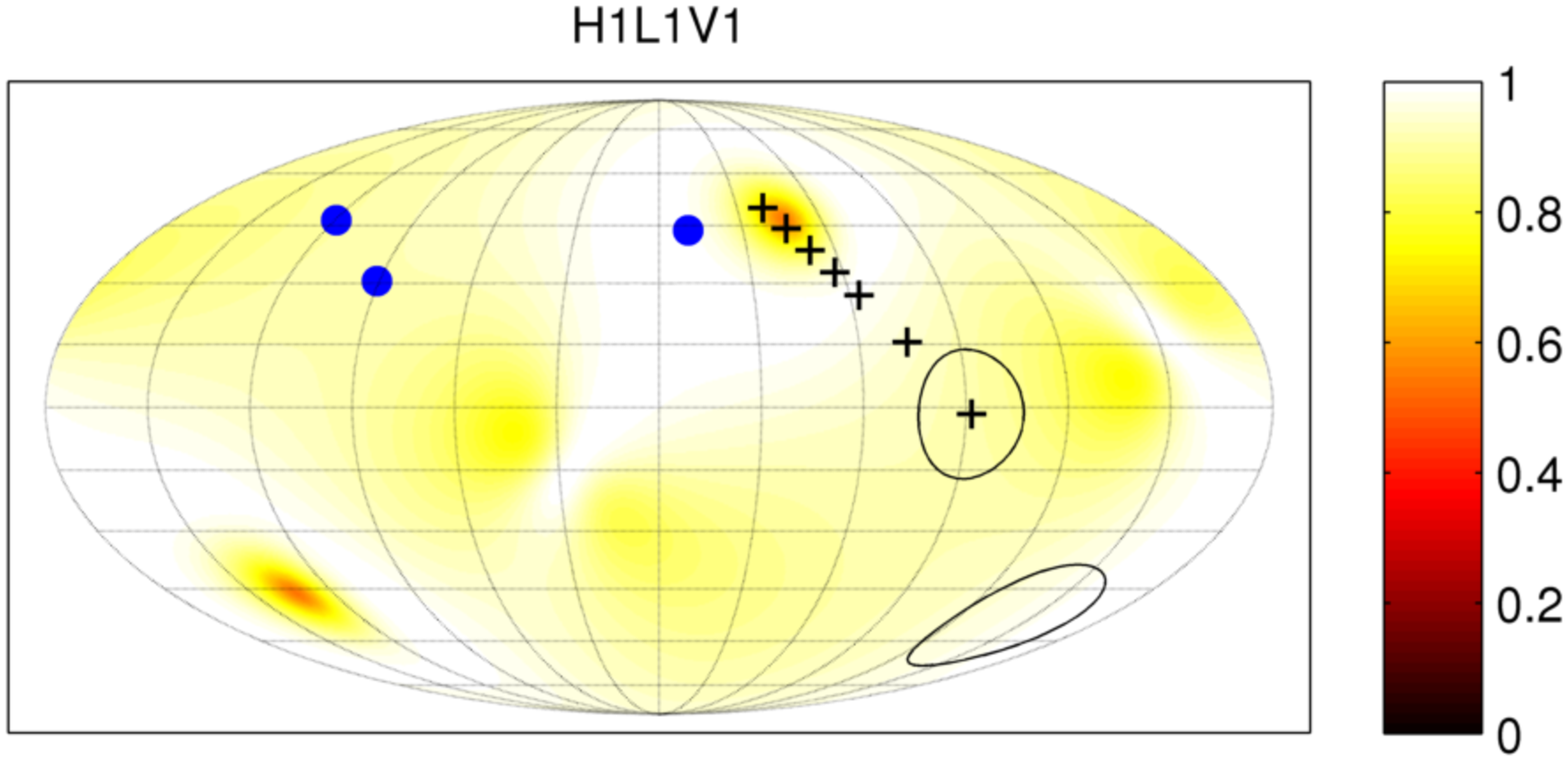}
  \caption{Sky dependence of the penalty factor $\lambda_c/\rho_c$ 
    [Eq.~\eqref{eq:penCohSNR2}] 
    assuming that V1 is a factor 2 less sensitive than H1 and L1.  From top to
    bottom the penalty factor for respectively the H1L1, L1V1 and
    H1L1V1 networks are shown. The
    coordinates used are the longitude and latitude in a Mollweid
    projection, as the detectors are fixed to the Earth. The location
    of the detectors in each network are marked by blue
    dots; for each detector the projection on the celestial sphere
    of the location (and its
    opposite) are the points of maximum antenna response.  
    For the 2 detector networks the zero points of the penalty factor 
    correspond to the null of the antenna response for one of the 
    two detectors.
    The plus marks show the sky location of the fiducial triggers
    analyzed to produce Fig.~\ref{fig:detectionFOM}, whereas the
    circles show the large sky regions used for the study shown in
    Fig.~\ref{fig:largeBoxFOM}. We omit the map for the H1V1 network
    as it is very similar to the L1V1 map. }
  \label{fig:penaltySky}
\end{figure}

To study the effect of different antenna pattern configurations, we
perform the analysis of fiducial external triggers well localized to
different points in the sky, which is representative of gamma-ray
bursts reported by Swift. We use the same time for all these
external triggers, but select sky locations which probe different 
combinations of contributions from the various detectors in the 
\gw\ detector network. The fiducial sky
locations and \gw\ detector networks used in this study are shown on the
3 panels of Fig.~\ref{fig:penaltySky}.

In order to have an \emph{a priori} measure of the signal strength
with regard to the detectors noise, we define the coherent network
SNR~\cite{Finn01}
\begin{equation}
  \rho_c = \sqrt{\sum_{\alpha \in \text{network}} \rho_\alpha^2}
\end{equation}
of a \gw\ signal model as the root-sum-square of the individual SNRs
\begin{equation}
\rho_\alpha^2  = 4 \int_0^\infty \frac{|F^+_\alpha h_+(f) +
  F^\times_\alpha h_\times(f)|^2}{S_\alpha(f)} \, \mathrm{d}f
\end{equation}
of that signal in each detector $\alpha$ across the
network. This figure of merit can be used as a predictor of the \gw\
signal amplitude (distance to the source) needed for the signal to be detected.

We define the {\it detection sensitivity} of the search for a particular signal model as
the distance $d_{50\%}$ at which that signal is found with $50\%$
efficiency while holding the background false-alarm probability fixed 
at $1\%$ per on-source window (in this case 1\% per 660 s, or a false 
alarm rate of $1.5\times10^{-5}$ Hz). 
This detection sensitivity is estimated by adding 
\gw\ signals into either real or simulated detector noise; 
approximately $10^4$ such ``injections'' are performed spread 
over 2 decades in distance. We use as our model signal a 
circularly polarized Gaussian-modulated sinusoid 
with central frequency $f_0 = 150$\,Hz and $Q=9$, 
\begin{multline}
  \begin{bmatrix}
    h_+(t) \\ h_\times(t) 
  \end{bmatrix} =
  \frac{A}{d}
  \begin{bmatrix}
    \cos(2 \pi f_0 t) \\
    \sin(2 \pi f_0 t)
  \end{bmatrix}\exp\left[ - \frac{(2 \pi f_0 t)^2}{2Q^2}\right] \, . \label{eqn:model}
\end{multline}
Here $A$ is an arbitrary scaling factor and $d$ the distance to the source.
This is a standard waveform for evaluating the sensitivity of triggered 
GWB searches \cite{burstGrbS5,grb051103,grbS6,2012arXiv1205.3018T,2008PhRvD..77f2004A}.

For an
ideal matched filter search in Gaussian noise $d_{50\%}$ should
correspond to the distance at which the median of $\rho_c$ for that
signal model crosses a certain threshold. This threshold depends on the 
number of degrees of freedom of the filtering template and the
effective number of independent times. 
For the search considered here, the number of independent
trials is of the order of $3 \times 10^5$: the total time-frequency 
volume in the on-source window.  The false alarm probability per trial 
is thus approximately $3 \times 10^{-8}$. The number of degrees of 
freedom is 2 (real and imaginary parts of the data) times the number 
of time-frequency pixels in a Gaussian noise cluster (typically in the 
4$-$6 range). Hence using a $\chi^2$ distribution for Gaussian noise we obtain an expected
threshold on $\rho_c$ in the 7.1$-$7.7 range.

However, for an analysis of real data, $\rho_c$ is not necessarily a
good figure of merit, as a signal needs also to pass coherent consistency 
tests to be distinguished from non-Gaussian noise transients. 
Heuristically, a signal needs a sufficient $E-I$
difference that Gaussian noise fluctuations will not destroy the signal
consistency as seen by the coherent tests described in
Sec.~\ref{sec:cohCuts}. Empirically we find that the penalized
coherent SNR
\begin{align}\label{eq:penCohSNR1}
  \lambda_c &= \rho_c \left[\frac{N_\text{det}+1}{N_\text{det}-1}
    \frac{E_\text{circular}^\text{signal} - I_\text{circular}^\text{signal}}{E^\text{signal}_\text{circular} +
      I_\text{circular}^\text{signal}} \right]^{1/4} \\
  &= \rho_c \left[ \frac{N_\text{det}+1}{N_\text{det}-1}
    \frac{\sum_{\alpha \neq \beta} |f_\alpha^\circlearrowright|^2
      |f_\beta^\circlearrowright|^2}{|\vb{f}^\circlearrowright|^4 +
      \sum_{\alpha} |f_\alpha^\circlearrowright|^4} \right]^{1/4}, \label{eq:penCohSNR2}
\end{align}
where $N_\text{det}$ is the number of detectors in the network, is a
good figure of merit for signals in non-Gaussian noise from real
\gw\ detectors \footnote{Equation (\ref{eq:penCohSNR2}) follows from 
(\ref{eq:penCohSNR1}) provided the relative noise levels $S_i(f)$ 
do not vary significantly across the bandwidth of the signal.}. 
The normalization factor
$(N_\text{det}+1)/(N_\text{det}-1)$ is chosen so that the penalty
factor $\lambda_c / \rho_c $ has a maximum value of 1. 
Note that the magnitudes of the right and left-handed sensitivity
vectors are equal
\begin{equation}
  |f_\alpha^\circlearrowright|=|f_\alpha^\circlearrowleft|= 
  \left({ f_\alpha^+}^2 + {f_\alpha^\times}^2 \right)^{1/2},
\end{equation}
so it does not matter which is used to evaluate
Eq.~\eqref{eq:penCohSNR2}.

The distribution on the sky of the penalty factor for different
networks is shown in Fig.~\ref{fig:penaltySky}. The maximum of the
factor for any network is attained when the sensitivity of all
detectors in the network is equal, that is when all
$|f_\alpha^\circlearrowright|$ are equal. For the 2 detector networks the
zero points of the penalty factor correspond to the blind spots of one
of the two detectors in the network.

\begin{figure}
  \centering
  \includegraphics[width=\linewidth]{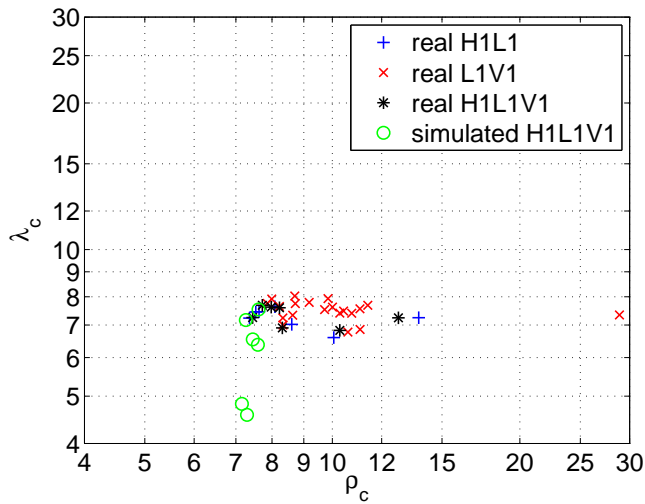} 
  \caption{Values of $\rho_c$ and $\lambda_c$ corresponding to the
    detection sensitivity $d_\text{50\%}$ found when analyzing real
    and simulated Gaussian noise. Each marker represents a complete 
    end-to-end analysis for one of the networks and sky positions 
    indicated in Fig.~\ref{fig:penaltySky}.
    Plus marks are for real noise from the
    H1L1 network, cross marks for real noise from the L1V1 network,
    star marks for real noise from the H1L1V1 network,
    and circles are for simulated Gaussian noise from the H1L1V1 network. The
    analyzed sky position are shown as crosses on
    Fig.~\ref{fig:penaltySky}.  }
  \label{fig:detectionFOM}
\end{figure}

The test sky positions used in the comparison are also shown in
Fig.~\ref{fig:penaltySky}. They are chosen along a line which
samples a wide range of penalty factors, in order to distinguish
$\lambda_c$ and $\rho_c$. We use a larger sample of test sky positions 
for the L1V1 network, as this is the network for which the penalty
factor effect is most important. 

For each of the trigger sky positions and networks indicated in 
Fig.~\ref{fig:penaltySky} we perform a complete analysis using 
both simulated Gaussian noise and real data.  In each case we 
compute the distance sensitivity $d_\text{50\%}$ at which the 
source model (\ref{eqn:model}) is detectable with 50\% probability 
at a fixed false alarm probability of 1\%. 
The values of $\rho_c$ and $\lambda_c$ corresponding to this $d_\text{50\%}$ 
are then computed for each test sky position; see Fig.~\ref{fig:detectionFOM}. 
For simulated noise we obtain the expected result that $d_\text{50\%}$
corresponds to a threshold on $\rho_c$, here equal to $\sim 7.4$,
which falls into the expected range of $7.1-7.7$.  However, for real
noise the obtained value of $\rho_c$ is spread over a much larger 
range of $7.5-30$, whereas $\lambda_c$ fluctuates by only 5\% around 
$\sim7.5$. Two conclusions can be drawn from these results:
\begin{enumerate}
\item $\lambda_c$ is a good predictor of the analysis sensitivity in
  real noise as a function of sky position for a given signal model;
\item  A sensitivity as good as in Gaussian noise ($\lambda_c\sim\rho_c$) 
  can be attained for sky positions where the penalty factor is close 
  to 1.  This occurs where all detectors have comparable sensitivity, 
  which corresponds to the white areas on Fig.~\ref{fig:penaltySky}.
\end{enumerate}

\subsection{Distribution of sensitive distance}\label{sec:sensDistrib}

The good performance of the penalized coherent SNR $\lambda_c$ in
predicting the sensitivity of real data analysis allows us to study
analytically the sensitivity sky dependence, and to compare it with
the Gaussian noise case which is given by $\rho_c$.

\begin{figure}
  \centering
  \includegraphics[width=\linewidth]{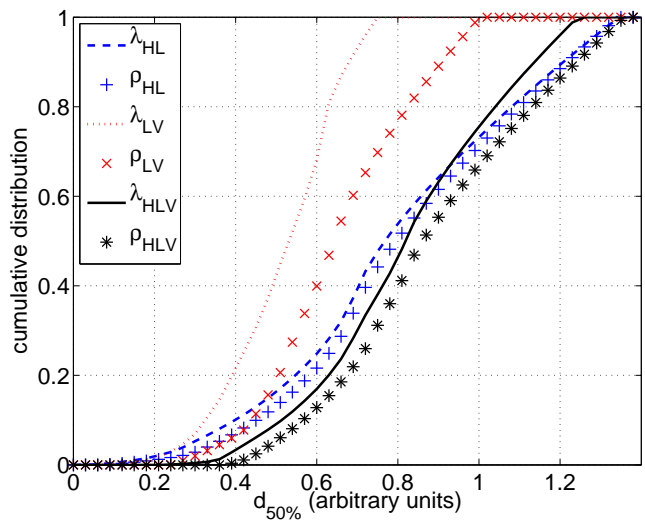} 
  \caption{Cumulative distribution over the sky of the sensitive distance
    $d_{50\%}$ assuming that V1 is a factor 2 less sensitive than
    H1 and L1. The lines show the distribution when $d_{50\%}$ is
    assumed to be given by a threshold on $\lambda_c$ (real data analysis): blue dashed
    line for the H1L1 network, red dotted line for the L1V1 network
    and black solid line for the H1L1V1 network. The marks show the
    distribution when $d_{50\%}$ is assumed to be given by a threshold
    on $\rho_c$ (Gaussian data analysis): blue pluses for the H1L1
    network, red crosses for L1V1 network and black stars for the
    H1L1V1 network.}
  \label{fig:penaltyDistrib}
\end{figure}

Fig.~\ref{fig:penaltyDistrib} shows the cumulative distribution of
the sensitive distance $d_{50\%}$ assuming that V1 is a factor 2 less
sensitive than H1 and L1. We consider two cases: a Gaussian noise
analysis where the sensitivity is given by a threshold on $\rho_c$ and
a real noise analysis where the sensitivity is predicted by the same
threshold but on $\lambda_c$.

For the H1L1 network the real and Gaussian noise
sensitivities are very close; our analysis of real data is only a few
percent less sensitive than the ideal Gaussian case. However for
networks including Virgo, especially for the L1V1 network of two
non-aligned detectors, the sensitivity is as much as 20\% lower with real
data than with ideal Gaussian noise. This is an expected effect of
each detector having maximum antenna response near the null
response of the other detector,
which limits the spurious transient noise rejection methods as they rely on
the signal being visible above the Gaussian part of the noise in at
least two detectors.

Interestingly for 30\% of the sky the sensitivity of the H1L1V1
network is slightly {\em worse} than that of the H1L1 network for the real data 
case.  The sensitivity loss occurs for
areas of the sky where the antenna patterns are optimal for the H1 and
L1 detectors, and relatively poor for V1. Hence for these sky regions
Virgo brings additional non-stationary noise but only a very small
increase in the total \gw\ signal. We note that the sensitivity 
difference is less than 10\%, and is a consequence of the
non-optimality of our analysis of real data (the optimal procedure 
for real data is not known). 
To verify this prediction, we repeat the full end-to-end 
analysis of a dozen sky positions in those regions for both the
H1L1 and H1L1V1 network. The comparison of the obtained $d_{50\%}$
confirms that the H1L1V1 network is less sensitive in those sky regions
by up to 10\%.  This indicates that care should be taken when selecting 
detectors for the analysis of real data in mis-aligned networks.

\subsection{Large sky area analysis}\label{sec:areaDep}

\begin{figure}
  \centering
  \includegraphics[width=\linewidth]{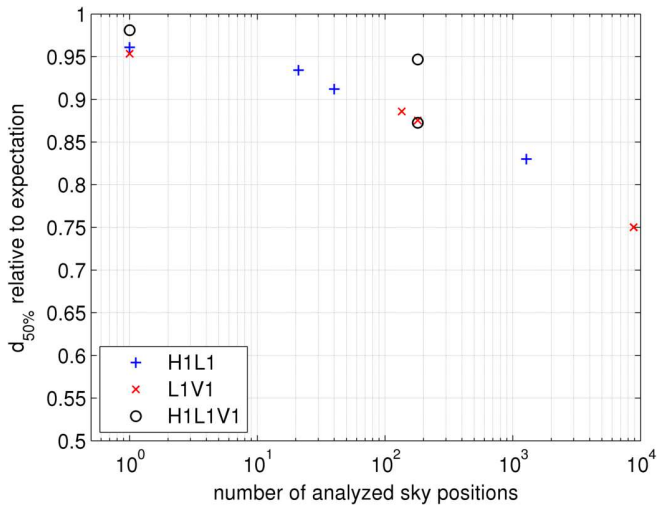}
  \caption{Fractional value of $d_\text{50\%}$ for real detector noise
    as a function of the number of analyzed sky positions relative to
    the expectation based on the typical threshold $\lambda_c = 7.5$,
    which we obtained in Fig.~\ref{fig:detectionFOM}. Blue plus marks
    denote results for the H1L1 network, red cross marks for the L1V1
    network and black circle marks for the H1L1V1 network. The 6 marks
    around 100 sky positions correspond to search sensitivity over the
    sky patches represented as circles on
    Fig.~\ref{fig:penaltySky}. The 2 marks with sky position number > 1000
    correspond to search sensitivity over the full sky.}
  \label{fig:largeBoxFOM}
\end{figure}

In principle, searching over a large sky area should lower the 
sensitivity, due to the trials factor incurred from repeating the 
analysis over a grid of sky positions.  We have seen that
the penalized coherent SNR $\lambda_c$ is a good figure of merit for
the sensitivity of \xp\ as a function of a single sky position.  
We therefore expect that for large sky areas 
the 50\% efficiency distance $d_{50\%}$ should correspond to a
threshold on the median value of $\lambda_c$ over the analyzed sky
area, where the median takes into account the prior
$p_\text{EM}(\Omega)$ on the true source sky position. 
Due to the trials factor, the threshold on this median $\lambda_c$ 
will typically be slightly higher than in the single
sky position case, and the ratio between the two yields an estimate of the
sensitivity loss due to a large sky area analysis.

To assess this sensitivity loss we repeat the analysis using typical
sky location uncertainties for the GBM instrument on Fermi: 
a $5^\circ$ statistical error and two-component systematic error 
as described in Ref.~\cite{GBMsysErr_GCN}. This results
in a search grid of $\sim$700 square degrees in area. Due to limited
computational resources only a small number of sky regions and 
network combinations are used; the analysed sky regions are shown
as circles on Fig.~\ref{fig:penaltySky}. For completeness, we also
perform a full-sky analysis for the H1L1 and L1V1 networks.

The resulting sensitive distances $d_\text{50\%}$ relative to the 
expected value for a single sky point analysis are shown in
Fig.~\ref{fig:largeBoxFOM}. We find that the performance loss of 
GBM-type error regions compared to precisely-localized external 
triggers is less than 10\%. A complete lack of sky 
position information (requiring an all-sky search) decreases the sensitivity 
by $\sim 20\%$. Hence the availability of
external triggers which are well localized on the sky ($\lesssim
1\unite{deg^2}$) improves the sensitivity by up to 20\%.
This is in addition to the sensitivity improvement resulting from 
the known time of the trigger, which reduces the trials factor from 
the length of data to be searched \cite{Kochanek93}.

\section{Conclusion}\label{sec:conclusion}

We have studied how the sensitivity of a search for GWBs performed by
\xp\ depends on the sky region specified by an external astrophysical
trigger. Two aspects of the sky region affect the search sensitivity: the 
magnitudes of the antenna patterns of the various \gw\ detectors in 
the network over this region; and the area of the region, which 
affects the size of the parameter space of the search. 

For the case of Gaussian background noise we have obtained the result
expected for an optimal analysis, namely the sensitivity is given by a
threshold on the coherent SNR $\rho_c$, and this threshold falls into
the range predicted by a $\chi^2$ distribution given the number of
independent trials in the search.

For real data we introduce a penalized coherent SNR $\lambda_c$, which
proves to be a good predictor of the search sensitivity in real non-Gaussian
noise. It is expressed as the coherent SNR times a penalty factor; this 
penalty factor is equal to 1 (no penalty) if all detectors have equal
sensitivity for a given sky position, and to 0 if only one of the
detectors is sensitive. For regions of the sky where the penalty
factor is equal to 1, the sensitivity in real noise is as good as in
the Gaussian noise case.

The penalized coherent SNR $\lambda_c$ allows us to separate the effect 
of antenna patterns changing over the sky from the effect of a 
search parameter space increase due to a large search sky region. 
We find that trigger information that allows us to restrict the search 
to a single
sky position increases the sensitivity by $\sim20\%$ compared to
searching over the whole sky, and by $\sim10\%$ compared to searching
over error boxes of a few hundred square degrees.  This addresses
part of the long-standing question on how externally triggered searches relate to
all-sky and all-time searches for \gws. 
%However, a full comparison
%requires discussing the astrophysics of a particular source, which is
%beyond the scope of this paper.

\begin{acknowledgments}
We thank Nicolas Leroy for valuable comments on an earlier draft of
this paper.  We thank the LIGO Scientific Collaboration for permission
to use data for our tests. LIGO was constructed by the California
Institute of Technology and Massachusetts Institute of Technology with
funding from the National Science Foundation and operates under
cooperative agreement number PHY-0107417. This paper has been assigned
LIGO document number LIGO-P1100135.
\end{acknowledgments}

\bibliography{References}

%Merlin.mbs v4.21 2009-07-09.
\begin{thebibliography}{10}%
\makeatletter
\providecommand \@ifxundefined [1]{%
 \ifx #1\undefined \expandafter \@firstoftwo
 \else \expandafter \@secondoftwo
\fi
}%
\providecommand \@ifnum [1]{%
 \ifnum #1\expandafter \@firstoftwo
 \else \expandafter \@secondoftwo
\fi
}%
\providecommand \enquote [1]{``#1''}%
\providecommand \bibnamefont  [1]{#1}%
\providecommand \bibfnamefont [1]{#1}%
\providecommand \citenamefont [1]{#1}%
\providecommand\href[0]{\@sanitize\@href}%
\providecommand\@href[1]{\endgroup\@@startlink{#1}\endgroup\@@href}%
\providecommand\@@href[1]{#1\@@endlink}%
\providecommand \@sanitize [0]{\begingroup\catcode`\&12\catcode`\#12\relax}%
\@ifxundefined \pdfoutput {\@firstoftwo}{%
 \@ifnum{\z@=\pdfoutput}{\@firstoftwo}{\@secondoftwo}%
}{%
 \providecommand\@@startlink[1]{\leavevmode\special{html:<a href="#1">}}%
 \providecommand\@@endlink[0]{\special{html:</a>}}%
}{%
 \providecommand\@@startlink[1]{%
  \leavevmode
  \pdfstartlink
   attr{/Border[0 0 1 ]/H/I/C[0 1 1]}%
   user{/Subtype/Link/A<</Type/Action/S/URI/URI(#1)>>}%
  \relax
 }%
 \providecommand\@@endlink[0]{\pdfendlink}%
}%
\providecommand \url  [0]{\begingroup\@sanitize \@url }%
\providecommand \@url [1]{\endgroup\@href {#1}{\urlprefix}}%
\providecommand \urlprefix [0]{URL }%
\providecommand \Eprint[0]{\href }%
\@ifxundefined \urlstyle {%
  \providecommand \doi [1]{doi:\discretionary{}{}{}#1}%
}{%
  \providecommand \doi [0]{doi:\discretionary{}{}{}\begingroup
  \urlstyle{rm}\Url }%
}%
\providecommand \doibase [0]{http://dx.doi.org/}%
\providecommand \Doi[1]{\href{\doibase#1}}%
\providecommand \bibAnnote [3]{%
  \BibitemShut{#1}%
  \begin{quotation}\noindent
    \textsc{Key:}\ #2\\\textsc{Annotation:}\ #3%
  \end{quotation}%
}%
\providecommand \bibAnnoteFile [2]{%
  \IfFileExists{#2}{\bibAnnote {#1} {#2} {\input{#2}}}{}%
}%
\providecommand \typeout [0]{\immediate \write \m@ne }%
\providecommand \selectlanguage [0]{\@gobble}%
\providecommand \bibinfo [0]{\@secondoftwo}%
\providecommand \bibfield [0]{\@secondoftwo}%
\providecommand \translation [1]{[#1]}%
\providecommand \BibitemOpen[0]{}%
\providecommand \bibitemStop [0]{}%
\providecommand \bibitemNoStop [0]{.\EOS\space}%
\providecommand \EOS [0]{\spacefactor3000\relax}%
\providecommand \BibitemShut [1]{\csname bibitem#1\endcsname}%
%</preamble>
\bibitem{burstS5y2}%
  \BibitemOpen
  \bibfield{author}{%
  \bibinfo {author} {\bibfnamefont{J.}~\bibnamefont{Abadie}} \emph{et~al.},\ }%
  \bibfield{journal}{%
  \bibinfo {journal} {Phys. Rev. D}\ }%
  \textbf{\bibinfo {volume} {81}},\ \bibinfo {pages} {102001} (\bibinfo {year}
  {2010})%
  \bibAnnoteFile{NoStop}{burstS5y2}%
\bibitem{burstGrbS5}%
  \BibitemOpen
  \bibfield{author}{%
  \bibinfo {author} {\bibfnamefont{B.~P.}\ \bibnamefont{Abbott}}
  \emph{et~al.},\ }%
  \bibfield{journal}{%
  \Doi{10.1088/0004-637X/715/2/1438}{\bibinfo {journal} {Astrophys. J.}}\ }%
  \textbf{\bibinfo {volume} {715}},\ \bibinfo {pages} {1438} (\bibinfo {year}
  {2010})%
  \bibAnnoteFile{NoStop}{burstGrbS5}%
\bibitem{Briggs09}%
  \BibitemOpen
  \bibfield{author}{%
  \bibinfo {author} {\bibfnamefont{M.}~\bibnamefont{Briggs}} \emph{et~al.},\ }%
  \bibfield{journal}{%
  \bibinfo {journal} {AIP Conf. Proc.}\ }%
  \textbf{\bibinfo {volume} {1133}},\ \bibinfo {pages} {40} (\bibinfo {year}
  {2009})%
  \bibAnnoteFile{NoStop}{Briggs09}%
\bibitem{GBMsysErr_GCN}%
  \BibitemOpen
  \bibfield{author}{%
  \bibinfo {author} {\bibfnamefont{V.}~\bibnamefont{Connaughton}},\ }%
  \bibinfo {howpublished} {{GCN circular 11574}} (\bibinfo {year} {2011})%
  \bibAnnoteFile{NoStop}{GBMsysErr_GCN}%
\bibitem{Baret:2011nu}%
  \BibitemOpen
  \bibfield{author}{%
  \bibinfo {author} {\bibfnamefont{B.}~\bibnamefont{Baret}} \emph{et~al.},\ }%
  \bibfield{journal}{%
  \Doi{10.1103/PhysRevD.85.103004}{\bibinfo {journal} {Phys. Rev. D}}\ }%
  \textbf{\bibinfo {volume} {85}},\ \bibinfo {pages} {103004} (\bibinfo {year}
  {2012})%
  \bibAnnoteFile{NoStop}{Baret:2011nu}%
\bibitem{2012arXiv1205.3018T}%
  \BibitemOpen
  \bibfield{author}{%
  \bibinfo {author}
  {\bibfnamefont{S.}~\bibnamefont{{Adri{\'a}n-Mart{\'{\i}}nez}}}
  \emph{et~al.}}%
   (\bibinfo {month} {May}\ \bibinfo {year} {2012}),\
  \Eprint{http://arxiv.org/abs/1205.3018}{arXiv:1205.3018}%
  \bibAnnoteFile{NoStop}{2012arXiv1205.3018T}%
\bibitem{Gursel89}%
  \BibitemOpen
  \bibfield{author}{%
  \bibinfo {author} {\bibfnamefont{Y.}~\bibnamefont{Gürsel}}\ and\ \bibinfo
  {author} {\bibfnamefont{M.}~\bibnamefont{Tinto}},\ }%
  \bibfield{journal}{%
  \Doi{10.1103/PhysRevD.40.3884}{\bibinfo {journal} {Phys. Rev. D}}\ }%
  \textbf{\bibinfo {volume} {40}},\ \bibinfo {pages} {3884} (\bibinfo {year}
  {1989})%
  \bibAnnoteFile{NoStop}{Gursel89}%
\bibitem{S4LIGOGEO}%
  \BibitemOpen
  \bibfield{author}{%
  \bibinfo {author} {\bibfnamefont{B.}~\bibnamefont{{Abbott}}} \emph{et~al.},\
  }%
  \bibfield{journal}{%
  \Doi{10.1088/0264-9381/25/24/245008}{\bibinfo {journal} {Classical and
  Quantum Gravity}}\ }%
  \textbf{\bibinfo {volume} {25}},\ \bibinfo {pages} {245008} (\bibinfo {month}
  {Dec.}\ \bibinfo {year} {2008})%
  \bibAnnoteFile{NoStop}{S4LIGOGEO}%
\bibitem{burstS5y1}%
  \BibitemOpen
  \bibfield{author}{%
  \bibinfo {author} {\bibfnamefont{B.~P.}\ \bibnamefont{Abbott}}
  \emph{et~al.},\ }%
  \bibfield{journal}{%
  \Doi{10.1103/PhysRevD.80.102001}{\bibinfo {journal} {Phys. Rev. D}}\ }%
  \textbf{\bibinfo {volume} {80}},\ \bibinfo {pages} {102001} (\bibinfo {year}
  {2009})%
  \bibAnnoteFile{NoStop}{burstS5y1}%
\bibitem{S5IMBBH}%
  \BibitemOpen
  \bibfield{author}{%
  \bibinfo {author} {\bibfnamefont{J.}~\bibnamefont{Abadie}} \emph{et~al.}}%
   (\bibinfo {month} {Jan.}\ \bibinfo {year} {2012}),\
  \Eprint{http://arxiv.org/abs/1201.5999}{arXiv:1201.5999}%
  \bibAnnoteFile{NoStop}{S5IMBBH}%
\bibitem{burstS6allsky}%
  \BibitemOpen
  \bibfield{author}{%
  \bibinfo {author} {\bibfnamefont{J.}~\bibnamefont{Abadie}} \emph{et~al.}}%
   (\bibinfo {month} {Feb.}\ \bibinfo {year} {2012}),\
  \Eprint{http://arxiv.org/abs/1202.2788}{arXiv:1202.2788}%
  \bibAnnoteFile{NoStop}{burstS6allsky}%
\bibitem{grb051103}%
  \BibitemOpen
  \bibfield{author}{%
  \bibinfo {author} {\bibfnamefont{J.}~\bibnamefont{Abadie}} \emph{et~al.}}%
   (\bibinfo {year} {2012}),\
  \Eprint{http://arxiv.org/abs/1201.4413}{arXiv:1201.4413}%
  \bibAnnoteFile{NoStop}{grb051103}%
\bibitem{grbS6}%
  \BibitemOpen
  \bibfield{author}{%
  \bibinfo {author} {\bibfnamefont{J.}~\bibnamefont{Abadie}} \emph{et~al.}}%
   (\bibinfo {year} {2012}),\
  \Eprint{http://arxiv.org/abs/1205.2216}{arXiv:1205.2216}%
  \bibAnnoteFile{NoStop}{grbS6}%
\bibitem{Note1}%
  \BibitemOpen
  \bibinfo {note} {For simplicity we consider only the case of a network
  composed of non-aligned detectors, as in the most recent science run of the
  LIGO-Virgo detector network (only one gravitational wave detector was
  operational at the LIGO-Hanford site during the 2009-2010 run).}%
  \bibAnnoteFile{Stop}{Note1}%
\bibitem{Klimenko05}%
  \BibitemOpen
  \bibfield{author}{%
  \bibinfo {author} {\bibfnamefont{S.}~\bibnamefont{Klimenko}}, \bibinfo
  {author} {\bibfnamefont{S.}~\bibnamefont{Mohanty}}, \bibinfo {author}
  {\bibfnamefont{M.}~\bibnamefont{Rakhmanov}},\ and\ \bibinfo {author}
  {\bibfnamefont{G.}~\bibnamefont{Mitselmakher}},\ }%
  \bibfield{journal}{%
  \Doi{10.1103/PhysRevD.72.122002}{\bibinfo {journal} {Phys. Rev. D}}\ }%
  \textbf{\bibinfo {volume} {72}},\ \bibinfo {pages} {122002} (\bibinfo {year}
  {2005})%
  \bibAnnoteFile{NoStop}{Klimenko05}%
\bibitem{Sylvestre02}%
  \BibitemOpen
  \bibfield{author}{%
  \bibinfo {author} {\bibfnamefont{J.}~\bibnamefont{Sylvestre}},\ }%
  \bibfield{journal}{%
  \Doi{10.1103/PhysRevD.66.102004}{\bibinfo {journal} {Phys. Rev. D}}\ }%
  \textbf{\bibinfo {volume} {66}},\ \bibinfo {pages} {102004} (\bibinfo {year}
  {2002})%
  \bibAnnoteFile{NoStop}{Sylvestre02}%
\bibitem{Searle08}%
  \BibitemOpen
  \bibfield{author}{%
  \bibinfo {author} {\bibfnamefont{A.~C.}\ \bibnamefont{Searle}}, \bibinfo
  {author} {\bibfnamefont{P.~J.}\ \bibnamefont{Sutton}}, \bibinfo {author}
  {\bibfnamefont{M.}~\bibnamefont{Tinto}},\ and\ \bibinfo {author}
  {\bibfnamefont{G.}~\bibnamefont{Woan}},\ }%
  \bibfield{journal}{%
  \Doi{10.1088/0264-9381/25/11/114038}{\bibinfo {journal} {Class. Quantum
  Grav.}}\ }%
  \textbf{\bibinfo {volume} {25}},\ \bibinfo {pages} {114038} (\bibinfo {year}
  {2008})%
  \bibAnnoteFile{NoStop}{Searle08}%
\bibitem{BAT05}%
  \BibitemOpen
  \bibfield{author}{%
  \bibinfo {author} {\bibfnamefont{S.~D.}\ \bibnamefont{Barthelmy}}
  \emph{et~al.},\ }%
  \bibfield{journal}{%
  \bibinfo {journal} {Space Sci. Rev.}\ }%
  \textbf{\bibinfo {volume} {120}},\ \bibinfo {pages} {143} (\bibinfo {year}
  {2005})%
  \bibAnnoteFile{NoStop}{BAT05}%
\bibitem{Briggs99}%
  \BibitemOpen
  \bibfield{author}{%
  \bibinfo {author} {\bibfnamefont{M.}~\bibnamefont{Briggs}} \emph{et~al.},\ }%
  \bibfield{journal}{%
  \bibinfo {journal} {Astrophys. J. Suppl. Ser.}\ }%
  \textbf{\bibinfo {volume} {122}},\ \bibinfo {pages} {503} (\bibinfo {year}
  {1999})%
  \bibAnnoteFile{NoStop}{Briggs99}%
\bibitem{IPN09}%
  \BibitemOpen
  \bibfield{author}{%
  \bibinfo {author} {\bibfnamefont{K.}~\bibnamefont{Hurley}} \emph{et~al.},\ }%
  \bibfield{journal}{%
  \bibinfo {journal} {AIP Conf. Proc}\ }%
  \textbf{\bibinfo {volume} {1133}},\ \bibinfo {pages} {55} (\bibinfo {year}
  {2009})%
  \bibAnnoteFile{NoStop}{IPN09}%
\bibitem{Cadonati04}%
  \BibitemOpen
  \bibfield{author}{%
  \bibinfo {author} {\bibfnamefont{L.}~\bibnamefont{Cadonati}},\ }%
  \bibfield{journal}{%
  \Doi{10.1088/0264-9381/21/20/012}{\bibinfo {journal} {Class. Quantum Grav.}}\
  }%
  \textbf{\bibinfo {volume} {21}},\ \bibinfo {pages} {S1695} (\bibinfo {year}
  {2004})%
  \bibAnnoteFile{NoStop}{Cadonati04}%
\bibitem{Wen05}%
  \BibitemOpen
  \bibfield{author}{%
  \bibinfo {author} {\bibfnamefont{L.}~\bibnamefont{Wen}}\ and\ \bibinfo
  {author} {\bibfnamefont{B.~F.}\ \bibnamefont{Schutz}},\ }%
  \bibfield{journal}{%
  \Doi{10.1088/0264-9381/22/18/S46}{\bibinfo {journal} {Class. Quantum Grav.}}\
  }%
  \textbf{\bibinfo {volume} {22}},\ \bibinfo {pages} {S1321} (\bibinfo {year}
  {2005})%
  \bibAnnoteFile{NoStop}{Wen05}%
\bibitem{Chatterji06}%
  \BibitemOpen
  \bibfield{author}{%
  \bibinfo {author} {\bibfnamefont{S.}~\bibnamefont{Chatterji}} \emph{et~al.},\
  }%
  \bibfield{journal}{%
  \Doi{10.1103/PhysRevD.74.082005}{\bibinfo {journal} {Phys. Rev. D}}\ }%
  \textbf{\bibinfo {volume} {74}},\ \bibinfo {pages} {082005} (\bibinfo {year}
  {2006})%
  \bibAnnoteFile{NoStop}{Chatterji06}%
\bibitem{Klimenko08}%
  \BibitemOpen
  \bibfield{author}{%
  \bibinfo {author} {\bibfnamefont{S.}~\bibnamefont{Klimenko}}, \bibinfo
  {author} {\bibfnamefont{I.}~\bibnamefont{Yakushin}}, \bibinfo {author}
  {\bibfnamefont{A.}~\bibnamefont{Mercer}},\ and\ \bibinfo {author}
  {\bibfnamefont{G.}~\bibnamefont{Mitselmakher}},\ }%
  \bibfield{journal}{%
  \Doi{10.1088/0264-9381/25/11/114029}{\bibinfo {journal} {Class. Quantum
  Grav.}}\ }%
  \textbf{\bibinfo {volume} {25}},\ \bibinfo {pages} {114029} (\bibinfo {year}
  {2008})%
  \bibAnnoteFile{NoStop}{Klimenko08}%
\bibitem{Sutton10}%
  \BibitemOpen
  \bibfield{author}{%
  \bibinfo {author} {\bibfnamefont{P.~J.}\ \bibnamefont{Sutton}}
  \emph{et~al.},\ }%
  \bibfield{journal}{%
  \Doi{10.1088/1367-2630/12/5/053034}{\bibinfo {journal} {New J. Phys.}}\ }%
  \textbf{\bibinfo {volume} {12}},\ \bibinfo {pages} {053034} (\bibinfo {year}
  {2010})%
  \bibAnnoteFile{NoStop}{Sutton10}%
\bibitem{Note2}%
  \BibitemOpen
  \bibinfo {note} {This happens when the orientation of the detector arms
  differs by $45^\circ $ when projected onto the plane orthogonal to the
  gravitational wave\ direction of propagation.}%
  \bibAnnoteFile{Stop}{Note2}%
\bibitem{Kobayashi03-1}%
  \BibitemOpen
  \bibfield{author}{%
  \bibinfo {author} {\bibfnamefont{S.}~\bibnamefont{Kobayashi}}\ and\ \bibinfo
  {author} {\bibfnamefont{P.}~\bibnamefont{Mészáros}},\ }%
  \bibfield{journal}{%
  \bibinfo {journal} {Astrophys. J. Lett.}\ }%
  \textbf{\bibinfo {volume} {585}},\ \bibinfo {pages} {L89} (\bibinfo {year}
  {2003})%
  \bibAnnoteFile{NoStop}{Kobayashi03-1}%
\bibitem{Shibata03}%
  \BibitemOpen
  \bibfield{author}{%
  \bibinfo {author} {\bibfnamefont{M.}~\bibnamefont{Shibata}}, \bibinfo
  {author} {\bibfnamefont{K.}~\bibnamefont{Shigeyuki}},\ and\ \bibinfo {author}
  {\bibfnamefont{E.}~\bibnamefont{Yoshiharu}},\ }%
  \bibfield{journal}{%
  \bibinfo {journal} {Mon. Not. R. Astron. Soc.}\ }%
  \textbf{\bibinfo {volume} {343}},\ \bibinfo {pages} {619} (\bibinfo {year}
  {2003})%
  \bibAnnoteFile{NoStop}{Shibata03}%
\bibitem{Davies02}%
  \BibitemOpen
  \bibfield{author}{%
  \bibinfo {author} {\bibfnamefont{M.~B.}\ \bibnamefont{Davies}}, \bibinfo
  {author} {\bibfnamefont{A.}~\bibnamefont{King}}, \bibinfo {author}
  {\bibfnamefont{S.}~\bibnamefont{Rosswog}},\ and\ \bibinfo {author}
  {\bibfnamefont{G.}~\bibnamefont{Wynn}},\ }%
  \bibfield{journal}{%
  \bibinfo {journal} {Astrophys. J. Lett.}\ }%
  \textbf{\bibinfo {volume} {579}},\ \bibinfo {pages} {L63} (\bibinfo {year}
  {2002})%
  \bibAnnoteFile{NoStop}{Davies02}%
\bibitem{Piro07}%
  \BibitemOpen
  \bibfield{author}{%
  \bibinfo {author} {\bibfnamefont{A.~L.}\ \bibnamefont{Piro}}\ and\ \bibinfo
  {author} {\bibfnamefont{E.}~\bibnamefont{Pfahl}},\ }%
  \bibfield{journal}{%
  \bibinfo {journal} {Astrophys. J.}\ }%
  \textbf{\bibinfo {volume} {658}},\ \bibinfo {pages} {1173} (\bibinfo {year}
  {2007})%
  \bibAnnoteFile{NoStop}{Piro07}%
\bibitem{Romero10}%
  \BibitemOpen
  \bibfield{author}{%
  \bibinfo {author} {\bibfnamefont{G.~E.}\ \bibnamefont{Romero}}, \bibinfo
  {author} {\bibfnamefont{M.~M.}\ \bibnamefont{Reynoso}},\ and\ \bibinfo
  {author} {\bibfnamefont{H.~R.}\ \bibnamefont{Christiansen}},\ }%
  \bibfield{journal}{%
  \bibinfo {journal} {Astron. Astrophys.}\ }%
  \textbf{\bibinfo {volume} {524}},\ \bibinfo {pages} {A4} (\bibinfo {year}
  {2010})%
  \bibAnnoteFile{NoStop}{Romero10}%
\bibitem{iLIGO}%
  \BibitemOpen
  \bibfield{author}{%
  \bibinfo {author} {\bibfnamefont{B.~P.}\ \bibnamefont{Abbott}}
  \emph{et~al.},\ }%
  \bibfield{journal}{%
  \Doi{10.1088/0034-4885/72/7/076901}{\bibinfo {journal} {Rep. Prog. Phys.}}\
  }%
  \textbf{\bibinfo {volume} {72}},\ \bibinfo {pages} {076901} (\bibinfo {year}
  {2009})%
  \bibAnnoteFile{NoStop}{iLIGO}%
\bibitem{Virgo}%
  \BibitemOpen
  \bibfield{author}{%
  \bibinfo {author} {\bibfnamefont{T.}~\bibnamefont{Accadia}} \emph{et~al.},\
  }%
  \bibfield{journal}{%
  \Doi{10.1088/1748-0221/7/03/P03012}{\bibinfo {journal} {JINST}}\ }%
  \textbf{\bibinfo {volume} {7}},\ \bibinfo {pages} {P03012} (\bibinfo {year}
  {2012})%
  \bibAnnoteFile{NoStop}{Virgo}%
\bibitem{Virgo11}%
  \BibitemOpen
  \bibfield{author}{%
  \bibinfo {author} {\bibfnamefont{T.}~\bibnamefont{Accadia}} \emph{et~al.},\
  }%
  \bibfield{journal}{%
  \Doi{10.1088/0264-9381/28/11/114002}{\bibinfo {journal} {Class. Quantum.
  Grav.}}\ }%
  \textbf{\bibinfo {volume} {28}},\ \bibinfo {pages} {114002} (\bibinfo {year}
  {2011})%
  \bibAnnoteFile{NoStop}{Virgo11}%
\bibitem{Finn01}%
  \BibitemOpen
  \bibfield{author}{%
  \bibinfo {author} {\bibfnamefont{L.~S.}\ \bibnamefont{Finn}},\ }%
  \bibfield{journal}{%
  \Doi{10.1103/PhysRevD.63.102001}{\bibinfo {journal} {Phys. Rev. D}}\ }%
  \textbf{\bibinfo {volume} {63}},\ \bibinfo {pages} {102001} (\bibinfo {year}
  {2001})%
  \bibAnnoteFile{NoStop}{Finn01}%
\bibitem{2008PhRvD..77f2004A}%
  \BibitemOpen
  \bibfield{author}{%
  \bibinfo {author} {\bibfnamefont{B.}~\bibnamefont{{Abbott}}} \emph{et~al.},\
  }%
  \bibfield{journal}{%
  \Doi{10.1103/PhysRevD.77.062004}{\bibinfo {journal} {\prd}}\ }%
  \textbf{\bibinfo {volume} {77}},\ \bibinfo {eid} {062004} (\bibinfo {month}
  {Mar.}\ \bibinfo {year} {2008})%
  \bibAnnoteFile{NoStop}{2008PhRvD..77f2004A}%
\bibitem{Note3}%
  \BibitemOpen
  \bibinfo {note} {Equation (\ref {eq:penCohSNR2}) follows from (\ref
  {eq:penCohSNR1}) provided the relative noise levels $S_i(f)$ do not vary
  significantly across the bandwidth of the signal.}%
  \bibAnnoteFile{Stop}{Note3}%
\bibitem{Kochanek93}%
  \BibitemOpen
  \bibfield{author}{%
  \bibinfo {author} {\bibfnamefont{C.~S.}\ \bibnamefont{Kochanek}}\ and\
  \bibinfo {author} {\bibfnamefont{T.}~\bibnamefont{Piran}},\ }%
  \bibfield{journal}{%
  \Doi{10.1086/187083}{\bibinfo {journal} {Astrophys. J. Lett.}}\ }%
  \textbf{\bibinfo {volume} {417}},\ \bibinfo {pages} {17} (\bibinfo {year}
  {1993})%
  \bibAnnoteFile{NoStop}{Kochanek93}%
\end{thebibliography}%

\end{document}